\documentclass[12pt]{reportj}
\usepackage{deluxetablej}
\usepackage{hyperref} 
\makeatletter
\def\@to{to}
\makeatother
\usepackage{times}
\usepackage{graphicx}
\usepackage{tocloft}
\usepackage{amssymb}
\usepackage{fancyheadings}
\usepackage{amsmath}
\emergencystretch=\maxdimen
\renewcommand\thesection{\arabic{section}}
\renewcommand\thesubsection{\thesection.\arabic{subsection}}

\def\ssection#1{\setcounter{subsection}{0} \refstepcounter{section} \section*{\hbox to \hsize{\large\bf \arabic{section}. #1\hfill }}\label{sec} \addcontentsline{toc}{section}{\arabic{section}. #1}}
\def\ssubsection#1{\setcounter{subsubsection}{0} \refstepcounter{subsection}\subsection*{\hbox to \hsize{\normalsize\bfseries\itshape \arabic{section}.\arabic{subsection} #1\hfill}}\label{subsec} \addcontentsline{toc}{subsection}{\arabic{section}.\arabic{subsection} #1}}
\def\ssubsubsection#1{\refstepcounter{subsubsection}\subsection*{\hbox to \hsize{\normalsize\bfseries\itshape \arabic{section}.\arabic{subsection}.\arabic{subsubsection} #1\hfill}}\label{subsubsec} \addcontentsline{toc}{subsubsection}{\arabic{section}.\arabic{subsection}.\arabic{subsubsection} #1}}

\def\ssectionstar#1{\section*{\hbox to \hsize{\large\bf #1\hfill}} \addcontentsline{toc}{section}{#1}}
\def\ssubsectionstar#1{\subsection*{\hbox to \hsize{\normalsize\bfseries\itshape #1\hfill}} \addcontentsline{toc}{subsection}{#1}}
\def\ssubsubsectionstar#1{\subsection*{\hbox to \hsize{\normalsize\it  #1\hfill}} \addcontentsline{toc}{subsection}{#1}}

\renewcommand{\cftaftertoctitle}{%
\mbox{}\hfill{\normalfont Page}}
\newcommand{\bdstar}{BD+28$^{\circ}$~4211}

\begin{document}

~\\

\vspace{-2.4cm}
\noindent\includegraphics*[width=0.295\linewidth]{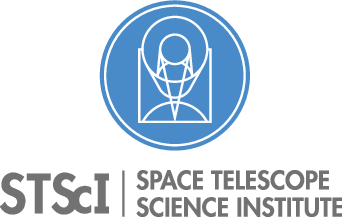}

\vspace{-0.4cm}

\begin{flushright}
    {\bf Instrument Science Report STIS 2024-04}
    
    \vspace{1.1cm}
    
    {\bf\Huge Updating the Sensitivity Curves of the STIS Echelles (Post-SM4)}
    
    \rule{0.25\linewidth}{0.5pt}
    
    \vspace{0.5cm}
    
    Svea Hernandez$^1$, TalaWanda Monroe$^1$, Joleen K. Carlberg$^1$
    \linebreak
    \newline
    \footnotesize{$^1$ Space Telescope Science Institute, Baltimore, MD\\}
    
    \vspace{0.5cm}
    
     21 August 2024
\end{flushright}

\vspace{0.1cm}

\noindent\rule{\linewidth}{1.0pt}
\noindent{\bf A{\footnotesize BSTRACT}}

{\it \noindent The STIS team re-derived on-orbit sensitivity curves for the echelle modes for post-servicing mission 4 observations using the standard DA white dwarf G 191-B2B. These new updates relied on the recent CALSPECv11 models, which introduced improvements in the fluxes of the primary standard stars of the order of $\sim$1-3\% depending on the wavelength of interest. As part of this effort, the team also released new blaze shift coefficients and echelle ripple tables. We present a detailed description of the procedures followed in the derivation of these new throughputs and the accompanying updates.}

\vspace{-0.1cm}
\noindent\rule{\linewidth}{1.0pt}

\renewcommand{\cftaftertoctitle}{\thispagestyle{fancy}}
\tableofcontents


\vspace{-0.3cm}
\ssection{Introduction}\label{sec:Introduction}
The Space Telescope Imaging Spectrograph (STIS) on board the Hubble Space Telescope (HST) provides users with access to four echelle grating modes with spectroscopic coverage between $\sim$1145 and 3100 \r{A}. These versatile gratings can be used as both medium- and high-resolution modes with resolving powers of $R\sim$25,000-45,000 ($\sim$10 km s$^{-1}$) and $R\sim$110,000 ($\sim$2.5 km s$^{-1}$), respectively. A detailed description of the individual echelle gratings can be found in the \href{https://hst-docs.stsci.edu/stisihb/chapter-13-spectroscopic-reference-material}{STIS Instrument Handbook Chapter 13}. \par
All of the echelle modes (primary and secondary) were fully calibrated back in 2006 after the instrument failed in 2004 (pre-Servicing Mission 4; \href{https://www.stsci.edu/files/live/sites/www/files/home/hst/instrumentation/stis/documentation/instrument-science-reports/_documents/200701.pdf}{Aloisi et al. 2007}). Once STIS was back in operations after Servicing Mission 4 (post-SM4) in May 2009, the team updated the sensitivity calibration of the echelle modes to account for any instrument changes including time and position dependent shifts in their blaze functions (\href{https://www.stsci.edu/files/live/sites/www/files/home/hst/instrumentation/stis/documentation/instrument-science-reports/_documents/2012_01.pdf}{Bostroem et al. 2012}). The post-SM4 sensitivity updates relied on the observations taken as part of the calibration program 11866 during Cycle 17, which collected observations with all STIS echelle modes. Generally, the derivation of sensitivity curves for a given instrument and mode requires the usage of a calibrator to define the absolute flux at a given wavelength. The calibrations by \href{https://www.stsci.edu/files/live/sites/www/files/home/hst/instrumentation/stis/documentation/instrument-science-reports/_documents/2012_01.pdf}{Bostroem et al. (2012)} made used of the pure hydrogen model for the standard G 191-B2B included in the CALSPECv07 library\footnote{\href{https://www.stsci.edu/hst/instrumentation/reference-data-for-calibration-and-tools/astronomical-catalogs/calspec}{https://www.stsci.edu/hst/instrumentation/reference-data-for-calibration-and-tools/astronomical-catalogs/calspec}} (\texttt{g191b2b\_mod\_007.fits}). \par
In 2020, \href{https://ui.adsabs.harvard.edu/abs/2020AJ....160...21B/abstract}{Bohlin et al.\ (2020)} introduced updated stellar atmospheric models for the three primary standard stars (GD 71, GD 153, and G 191-B2B) used to flux calibrate all of the HST instruments, as well as a re-examination of the Vega spectral flux, similarly impacting the absolute fluxes of these same standards. These model improvements were captured under the CALSPECv11 database. We note that one of the major changes in the CALSPECv11 library, compared to that from CALSPECv07, involved the inclusion of a metal-line-blanketed model atmosphere for G 191-B2B which redistributed its output flux as a function of wavelength. The combined changes in the spectral energy distributions (SEDs) of the three primary standards required the flux recalibration of all of the HST instruments, including STIS. \par

STIS offers a total of 44 different echelle settings across four different gratings: E140M, E140H, E230M, and E230H. As part of this ambitious flux recalibration effort, the STIS team prioritized the most widely used configurations. In this instrument science report (ISR) we present a detailed account of the derivation of the sensitivity curves, and other calibration features, using the CALSPECv11 models. We highlight that the changes to the flux calibration of the prioritized echelle settings described here apply only to data taken after SM4.  
Given the monthly offsets applied in the spatial and spectral direction to echelle spectra prior to SM4, a different strategy was employed for pre-SM4 era data. The details of this method are described in a separate ISR (\href{https://www.stsci.edu/files/live/sites/www/files/home/hst/instrumentation/stis/documentation/instrument-science-reports/_documents/STIS_ISR_2024-02.pdf}{Siebert et al., 2024}).

\lhead{}
\rhead{}
\cfoot{\rm {\hspace{-1.9cm} Instrument Science Report STIS 2024-04(v1) Page \thepage}}

\vspace{-0.3cm}
\ssection{Observations}\label{sec:obs}
\ssubsection{Archival PID 11866}\label{11866}
Similar to the early post-SM4 flux calibration, the re-derivation of the sensitivity curves for the echelle gratings was done through the analysis of the data taken as part of the Cycle 17 program 11866. This calibration program collected observations of the HST primary standard WD G 191-B2B between November 2009 and January 2010 with every echelle observing mode. Generally, the chosen exposure times targeted signal-to-noise (S/N) ratios between 20 and 100, depending on the configuration (see \href{https://www.stsci.edu/files/live/sites/www/files/home/hst/instrumentation/stis/documentation/instrument-science-reports/_documents/2012_01.pdf}{Bostroem et al. 2012} for more details on the observations from this program). \par

\ssubsection{Archival PID 15381: E140M}\label{15381}
The special calibration program 15381 was executed in February 2018 observing the primary standard star G 191-B2B in an effort to investigate changes in the blaze function shapes of the E140M/1425 setting that could not be accounted for with the blaze shift coefficients alone. These observations were previously used to re-characterize the blaze function shape of E140M/1425 and re-derive sensitivity curves (e.g., \href{https://www.stsci.edu/files/live/sites/www/files/home/hst/instrumentation/stis/documentation/instrument-science-reports/_documents/2022-04.pdf}{Carlberg et al. 2022}).

\begin{table}[!h] 
  \centering
    \caption{Exposure information for the datasets used in updating the prioritized echelle settings.}\label{tab:data}
    \def\arraystretch{1.25}
    \begin{tabular}{c c c c c}
    \hline
    \hline
   Grating & CENWAVE & Program ID & Dataset & Exposure Time \\
   & & & & (s)\\
    \hline
    E140M$^{1}$ & 1425 & 11866  & obb004070 & 695 \\
  &  & &  obb0040a0 & 642 \\
 &   & &  obb0040b0 & 3200\\
  &  & 15381 & odqw01010 & 703 ($\times$3) \\
    E230M & 1978  &  11866 & obb004050 & 455 \\
       & 2415 & 11866 & obb004040 &  280\\
     & 2707  &  11866 & obb004060 & 345\\
    E230H & 2263 & 11866 & obb002060 & 800  \\
    & 2713 & 11866 & obb053050 & 900  \\
    \hline
    \end{tabular}
    \\ $^{1}$ Previously reported in \href{https://www.stsci.edu/files/live/sites/www/files/home/hst/instrumentation/stis/documentation/instrument-science-reports/_documents/2022-04.pdf}{Carlberg et al. (2022)}.
\end{table}

\vspace{-0.3cm}
\ssection{Flux Recalibration Updates}\label{sec:sens_curve}
Previous to the STIS flux recalibration work driven by the new CALSPECv11 updates, the sensitivities and blaze shift coefficients of the E140M/1425 setting were last updated in 2020 (\href{https://www.stsci.edu/contents/news/stis-stans/february-2020-stan.html#e140m}{February 2020 STAN}). \href{https://www.stsci.edu/files/live/sites/www/files/home/hst/instrumentation/stis/documentation/instrument-science-reports/_documents/2022-04.pdf}{Carlberg et al. (2022)} describe in detail the recalibration steps followed as part of that work, 
and the work presented here follows a similar approach in creating updated photometric throughput files (PHOTTAB) and ripple tables (RIPTAB). Given that the blaze function of the echelle gratings is known to shift over time and depend on detector location, its characterization is used in both the derivation of the sensitivity curves and in the 2-dimensional (2D) echelle scattered light background subtraction derived from the RIPTABs. We note that the throughput files are created iteratively following the update of the ripple functions given that the latter impacts the extraction of the NET count rate used to derive the sensitivities themselves. This in turn means that the PHOTTAB and RIPTAB need to be updated simultaneously. Overall, the recalibration approach can be briefly described in four major steps: 1) derivation of sensitivity curves, 2) derivation of the ripple function, 3) derivation of the time-dependent blaze shift coefficients, and 4) conversion of the normalized sensitivities to efficiency units (throughputs). 

\ssubsection{Sensitivity Curves}\label{sens_curves}
The sensitivity curves, S($\lambda$), are derived for individual settings (grating and central wavelength), 
generally, by dividing the NET counts of a given standard star exposure (in this case of G 191-B2B), by a well calibrated model atmosphere (CALSPECv11). For this update, the STIS observations were calibrated with \texttt{CALSTIS v3.4.2}, and the NET counts and wavelength arrays extracted from the resulting ``x1d" files. The observed spectrum is then divided by the G 191-B2B model SED (\texttt{g191b2b\_mod\_011.fits}) interpolated to the wavelength of the observed spectrum. S($\lambda$) is defined in the observed reference frame, meaning that the heliocentric and doppler corrections applied by \texttt{CALSTIS} are removed from the individual STIS exposures. We note that in contrast to the previously used CALSPECv07 model, the updated CALSPECv11 SED already accounts for the radial velocity of G 191-B2B (22.0 km s$^{-1}$; \href{https://ui.adsabs.harvard.edu/abs/1988ApJ...335..953R/abstract}{Reid \& Wegner   1988}). Similarly, the wavelength scale of the model SED was modified to match the observed reference frame of the STIS data.  \par
\begin{figure}[!]
  \centering
  \includegraphics[width=1\textwidth]{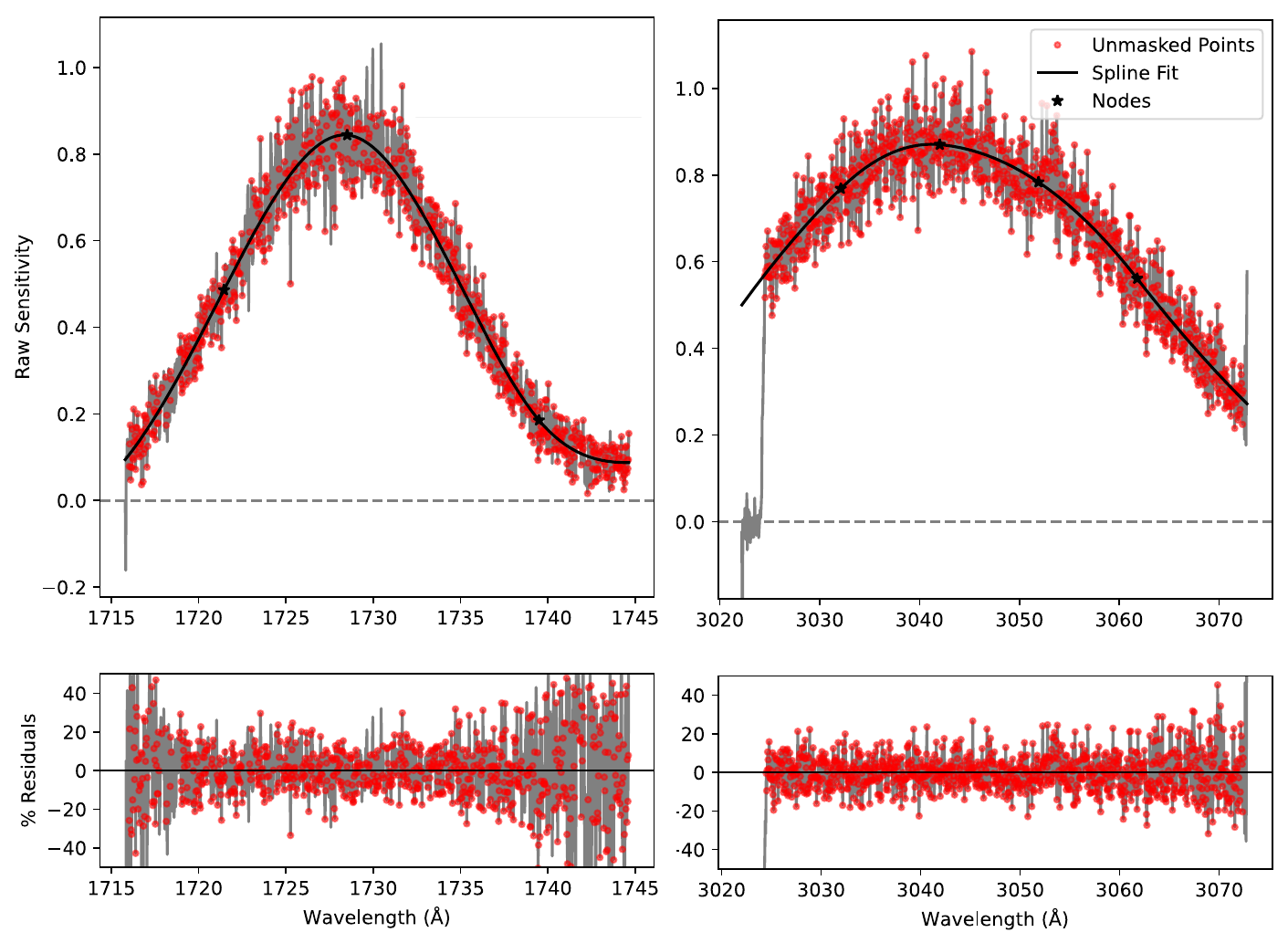} \\
    \caption{Derived sensitivity curves for example orders to showcase the shape of the different sensitivities across orders. We show in the bottom panels the percent residuals. \textit{Left:} Order 118 in E230M/1978 setting. \textit{Right:} Order 67 in the E230M/2707 setting. We highlight the masking of the points at the bluest wavelengths impacted by vignetting. Similar masking was applied to strong absorption lines and other detector features.}
    \label{fig:sens_e230m_2707}
\end{figure}

As stated earlier, the sensitivity curves were previously derived using the CALSPECv07 models, which did not include stellar absorption features (metal lines) other than the strong hydrogen Lyman series. To accurately derive the sensitivities, we identified wavelength regions affected by non-negligible stellar absorption and mask them accordingly. The recent CALSPECv11 G 191-B2B model includes an additional continuum array. This particular array was used only for identifying and removing the effects of line blanketing. In addition to masking strong stellar absorption lines, we also exclude strong ISM absorption features, unidentified stellar photospheric lines, the strong Lyman $\alpha$ and $\beta$ features, and strongly vignetted regions typically found on the edges of the orders (see right panel in Figure \ref{fig:sens_e230m_2707}). \par
The masked sensitivity curves for the individual orders are then fit multi-node quadratic splines. The location and number of nodes (between 2 and 5) varied between orders and configurations. These were chosen to identify the true and unbiased sensitivity curves. An example of this optimization is shown in the left panel in Figure \ref{fig:sens_e230m_2707}, where we strategically place the reddest node, at $\sim$ 1738 \r{A}, to better fit the curvature on the edge of the order. Similar to previous sensitivity derivations, when more than one observation was available for a given setting, we fit each exposure separately and the fits were later combined without weights. 

\ssubsection{Ripple Function (RIPTAB)}\label{riptab}
The resulting sensitivity fits were then evaluated onto the wavelength arrays for the corresponding orders stored in the ripple table (RIPTAB). These are then normalized to create the first updated version of the ripple functions for each spectral order. Briefly, the ripple functions are used to create 2-D scattered light models which are then subtracted from the calibrated FLT image. We highlight that the creation of the 2-D scattered light model within \texttt{CALSTIS} is done iteratively, where self-consistency between the model image and the science data is achieved after only three iterations (\href{https://www.stsci.edu/files/live/sites/www/files/home/hst/instrumentation/stis/documentation/instrument-science-reports/_documents/200201.pdf}{Valenti et al. 2002}). \par 
With the ripple functions at hand, the STIS observations are re-calibrated using \texttt{CALSTIS} and the updated RIPTAB, and setting the header keyword \texttt{FLUXCORR=OMIT} in the raw files. This second recalibration, with updated ripple functions generates slightly different NET count rates stored in the x1d files, which are then used in a second and final iteration of the steps described in section \ref{sens_curves} to obtain the optimal sensitivities and final version of the ripple table. 

\begin{figure}[!]
  \centering
  \includegraphics[width=1\textwidth]{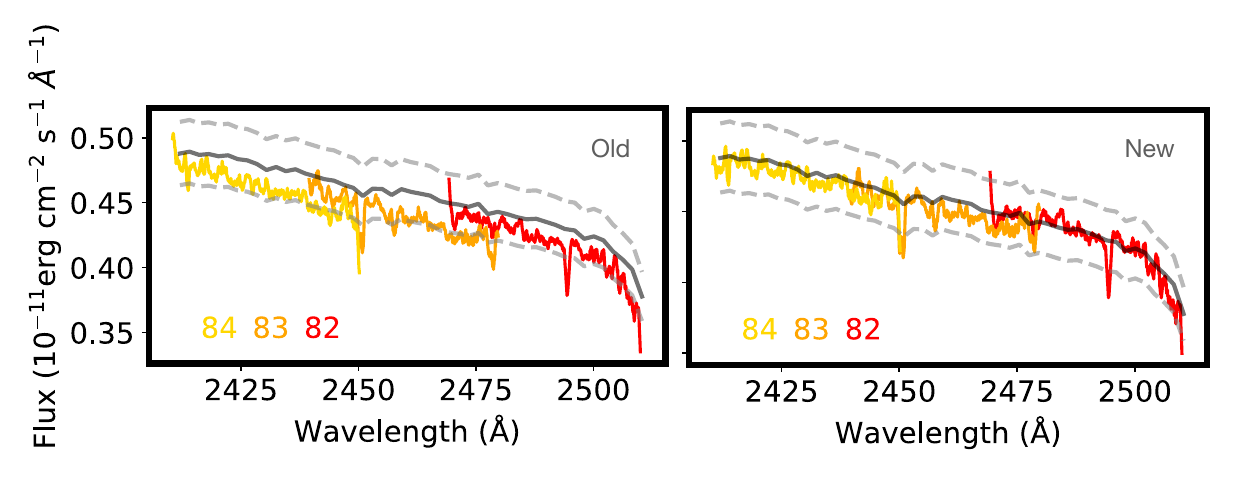} \\
    \caption{STIS E230M/2707 spectra for orders 82 (red), 83 (orange), and 84 (yellow). We show with a solid black curve the STIS CALSPECv11 spectrum (interpolated to the wavelength grid of the E230M/2707 data), and the $\pm$5 \% intervals in dashed lines. The left plot shows the previous calibration, contrasted to the right plot which shows the latest calibrations (new PHOTTAB, including updated blaze shift coefficients, and RIPTTAB).}
    \label{fig:blaze_e230m_2707}
\end{figure}

\ssubsection{Throughput conversion (PHOTTAB)}\label{phttab}
Given that the observations used to derive these updated sensitivities were taken after the reference time of MJD 50587.0 (or 1997.37 in decimal year), these need to be corrected for any time-dependent changes. These corrections are applied using the latest time dependent sensitivity table (TDSTAB). Lastly, these TDS-corrected, wavelength-dependent sensitivities are in units of (counts/s)/(ergs/cm$^{2}$/s/$\rm \r{A}$), which describe the telescope throughput as measured through a 0.2"$\times$0.2" aperture (configuration of the observations used to estimate these throughputs), and extracted using the nominal 7-pixel extraction box height. These sensitivities are then converted to throughput curves (or efficiencies for an infinite aperture and infinite extraction box) following equation (1) in  \href{https://www.stsci.edu/files/live/sites/www/files/home/hst/instrumentation/stis/documentation/instrument-science-reports/_documents/2012_01.pdf}{Bostroem et al.\ (2012)}. Briefly, this conversion involves aperture and extraction box corrections applied using standard reference files, as well as accounting for the collecting area of HST, storing the throughputs in updated versions of the PHOTTAB reference file.

\ssubsection{Recovery of echelle edge orders}\label{new_orders}
In the calibration by \href{https://www.stsci.edu/files/live/sites/www/files/home/hst/instrumentation/stis/documentation/instrument-science-reports/_documents/2012_01.pdf}{Bostroem et al. (2012)} for post-SM4 data, several spectral orders present in the observations of program 11866 were excluded from the PHOTTAB files, and therefore excluded from the final flux calibrated \texttt{CALSTIS} products. The spectral traces for these orders, however, remained in the spectral table reference file (SPTRCTAB) allowing us to derive updated throughputs by calibrating the observations only after setting the header keyword \texttt{FLUXCORR=OMIT}. Under such a setting, \texttt{CALSTIS} is able to extract the NET count rate of all of the orders present in the SPTRCTAB. In this latest calibration update, we once again include throughputs for some of these previously-excluded orders in the PHOTTABs and RIPTTABs allowing \texttt{CALSTIS} to fully calibrate these additional orders. In Table \ref{tab:new_orders} we list the newly flux calibrated orders and their corresponding STIS setting and wavelength coverage. We note that because these additional orders are located on the bottom and top edges of the detector, one of the two pre-defined background regions for each order falls outside of the detector.  The background contributions for these newly-added orders can then be underestimated, and the data quality (DQ) flag 2048 ($>$30\% of background pixels rejected by sigma-clipping) would be assigned to all events in that particular order. We highlight that the calibrated fluxes of high signal-to-noise observations are not strongly impacted by this issue.

\begin{table}[!h] 
  \centering
    \caption{Previously-excluded orders now available in the updated PHOTTAB and RIPTAB files}\label{tab:new_orders}
    \def\arraystretch{1.25}
    \begin{tabular}{l | c c c}
    \hline
    \hline
   Grating & CENWAVE & Order & Wavelength Coverage \\
   & & & ($\rm \r{A}$)\\
    \hline
    E140M$^{1}$ & 1425 & 86  & 1710--1730  \\
    E230M & 2707  &  66 & 3075--3125 \\
    & 2415 & 73 & 2780--2820 \\
    E230H & 2713 & 272 & 2830--2845  \\
    & & 300 & 2560--2580 \\
    \hline
    \end{tabular}
    \label{tab:table_one} 
    \\ $^{1}$ Previously reported in \href{https://www.stsci.edu/files/live/sites/www/files/home/hst/instrumentation/stis/documentation/instrument-science-reports/_documents/2022-04.pdf}{Carlberg et al. (2022)}.
\end{table}

\ssubsection{Blaze Shift Coefficients}\label{bs}

Updated characterizations of the spectral blaze functions for the E230M and E230H modes required changes to the temporal components of the blaze shift coefficients.  Over time the blaze functions can shift with respect to the wavelength scale (\href{https://www.stsci.edu/files/live/sites/www/files/home/hst/instrumentation/stis/documentation/instrument-science-reports/_documents/200701.pdf}{Aloisi et al. 2011}; \href{https://www.stsci.edu/files/live/sites/www/files/home/hst/instrumentation/stis/documentation/instrument-science-reports/_documents/2012_01.pdf}{Bostroem et al. 2012}; \href{https://www.stsci.edu/files/live/sites/www/files/home/hst/instrumentation/stis/documentation/instrument-science-reports/_documents/2022-04.pdf}{Carlberg et al. 2022}), which can result in flux mismatches in the overlapping spectral regions of adjacent orders of up to $\sim$10\%.  In the post-SM4 era, time dependent coefficients for the blaze function shifts were previously derived and updated for all cenwaves of the E230M grating in July 2019.  However, until now, no time dependent coefficients had been derived for post-SM4 E230H observations.

The functional form of the blaze shift equation is defined in \href{https://www.stsci.edu/files/live/sites/www/files/home/hst/instrumentation/stis/documentation/instrument-science-reports/_documents/200701.pdf}{Aloisi et al. (2007)} as:

\begin{multline}
BZS =  BSHIFT\_VS\_X \cdot \Delta x + BSHIFT\_VS\_Y \cdot \Delta y + BSHIFT\_VS\_T \cdot \Delta t \\  + BSHIFT\_OFFSET.
\label{eq:blaze_eqn}
\end{multline}

\noindent As part of the work described in this report our team derived  the \texttt{BSHIFT\_VS\_T} and \texttt{BSHIFT\_OFFSET} coefficients from equation \ref{eq:blaze_eqn} for the E230M and E230H gratings, which are presented in Table \ref{tab:blaze_coeff_tbl}. We note that the blaze shift coefficients previously derived for E140M are described in \href{https://www.stsci.edu/files/live/sites/www/files/home/hst/instrumentation/stis/documentation/instrument-science-reports/_documents/2022-04.pdf}{Carlberg et al. (2022)}.

The procedure for deriving blaze shift coefficients is given in detail in \href{https://www.stsci.edu/files/live/sites/www/files/home/hst/instrumentation/stis/documentation/instrument-science-reports/_documents/2022-04.pdf}{Carlberg et al. (2022)}, under Section 6.1, and will only be briefly summarized here.  Reference observations of \bdstar\ were used to derive the coefficients for the primary settings E230M/1978 and 2707, and E230H/2263 (e.g., PIDs 11860, 12414, and 12775).  These observations were chosen to be close in time to the G 191-B2B sensitivity curve observations, serving as a temporal reference point.  For each observation of  \bdstar, the NET count rates of each spectral order are fitted as a function of wavelength with a low order polynomial, which is then normalized to its maximum value. The blaze function shift is computed for each spectral order via cross correlation to the reference \bdstar\ observation for each mode.  The spatial components of the blaze shift are subtracted from the cross correlation values to obtain a residual temporal component shift. For each order, we apply a linear fit to the residual temporal shifts as a function of time with respect to the reference observation. The \texttt{BSHIFT\_VS\_T} and \texttt{BSHIFT\_OFFSET} coefficients are defined as the slope and $y$-intercept of the fits, respectively.  The groups of coefficients are then linearly fitted as a function of spectral order to reduce noise within the fit of each order.

\begin{figure}[!]
  \centering
  \includegraphics[width=1\textwidth]{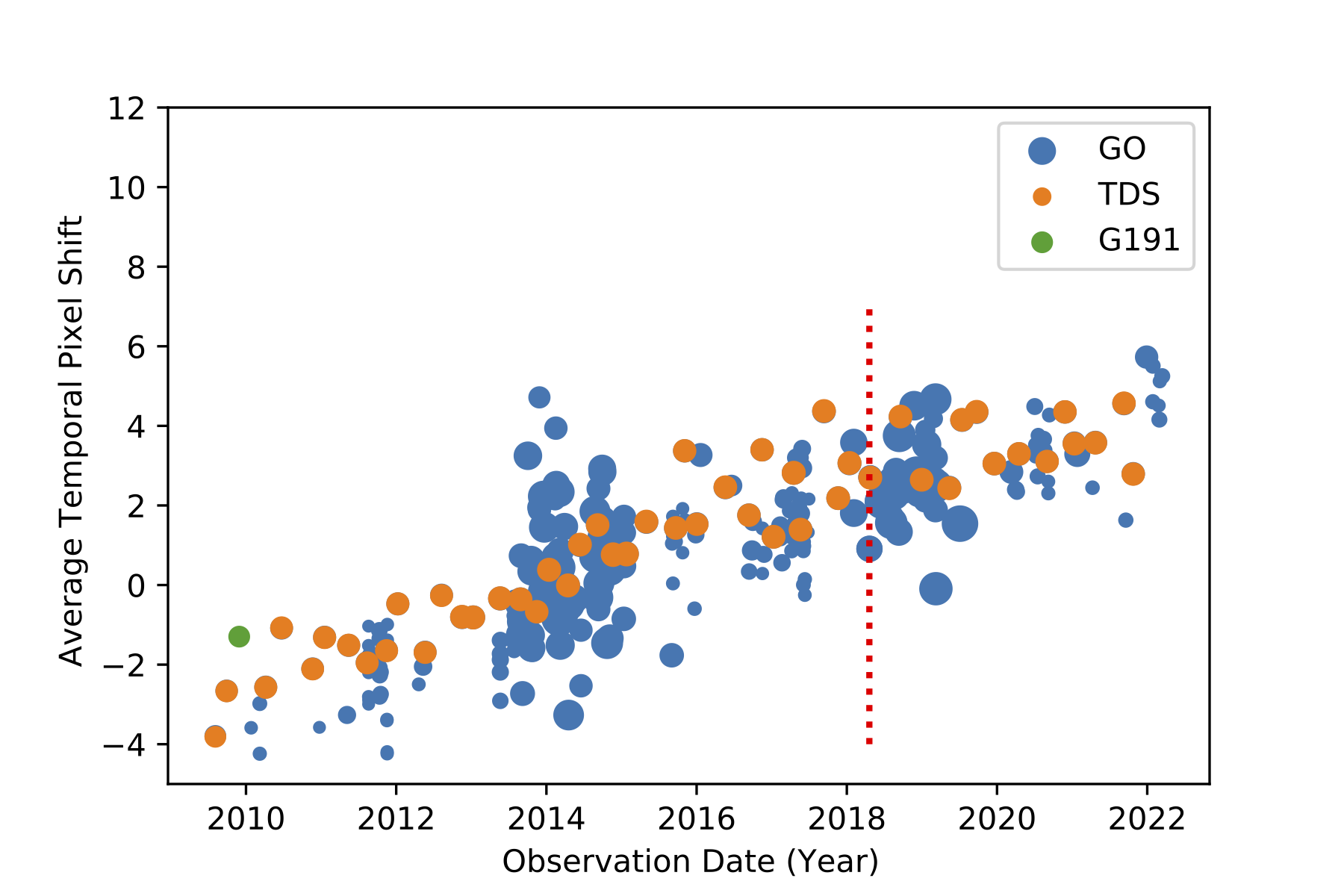} \\
    \caption{Average pixel shifts calculated by the \texttt{stisblazefix} tool for post-SM4 E230M/1978 observations with S/N ratios greater than 10 in the central spectral orders.  GO observations are indicated in blue, monitoring observations of \bdstar\ are indicated in orange, and the G 191-B2B reference observation is included in green.  A subtle inflection in the expected linear trend of the temporal blaze shifts is apparent $\sim$2018.3 (red vertical line).}
    \label{fig:E230M_1978_inflection}
\end{figure}

\ssubsubsection{E230M Coefficients}\label{e230m_bs}
Observations of \bdstar $\:$ taken with E230M central wavelengths of 1978 and 2707 were used to derive separate temporal blaze shift coefficients.  Given that both modes use the same grating, it was expected that the blaze shift time trends would produce similar  coefficient terms.  While the \texttt{BSHIFT\_VS\_T} slopes were in close agreement for both cenwaves, the optimal \texttt{BSHIFT\_OFFSET} constant terms were different for each cenwave.  Furthermore, it appeared there was a slight inflection in the expected linear trends of the settings occurring around decimal year 2018.3.  To explore whether this inflection was also present in GO observations, we used the standalone \texttt{stisblazefix} tool (\href{https://www.stsci.edu/files/live/sites/www/files/home/hst/instrumentation/stis/documentation/instrument-science-reports/_documents/2018_01.pdf}{Baer et al. 2018}) to compute blaze shifts for all post-SM4 archival observations that used the E230M/1978 and E230M/2707 settings.  Figure \ref{fig:E230M_1978_inflection} shows the temporal pixel shifts for post-SM4 E230M/1978 observations.  A subtle inflection in the linear trend is evident $\sim$2018.3 for the GO observations shown in blue and the monitoring observations of \bdstar\ shown in orange.  A similar inflection was found in the E230/2707 observations.  Observations after the inflection are better characterized by a shallower blaze shift fit. We note that due to the observed trends we adopted separate blaze shift coefficients for each central wavelength and each time period.  The format of the spectroscopic PHOTTAB file used with \texttt{CALSTIS} does not allow for a breakpoint in the coefficients, so two separate PHOTTAB files were delivered to CRDS with different USEAFTER dates (i.e., March 31, 2018, for the later set of coefficients).  Additionally, the REFMJD date for the E230M/1978 mode was manually changed for observations after March 31, 2018, to provide a better linear fit for the coefficients, as indicated in Table \ref{tab:blaze_coeff_tbl}.

\begin{figure}[!]
  \centering
  \includegraphics[width=1\textwidth]{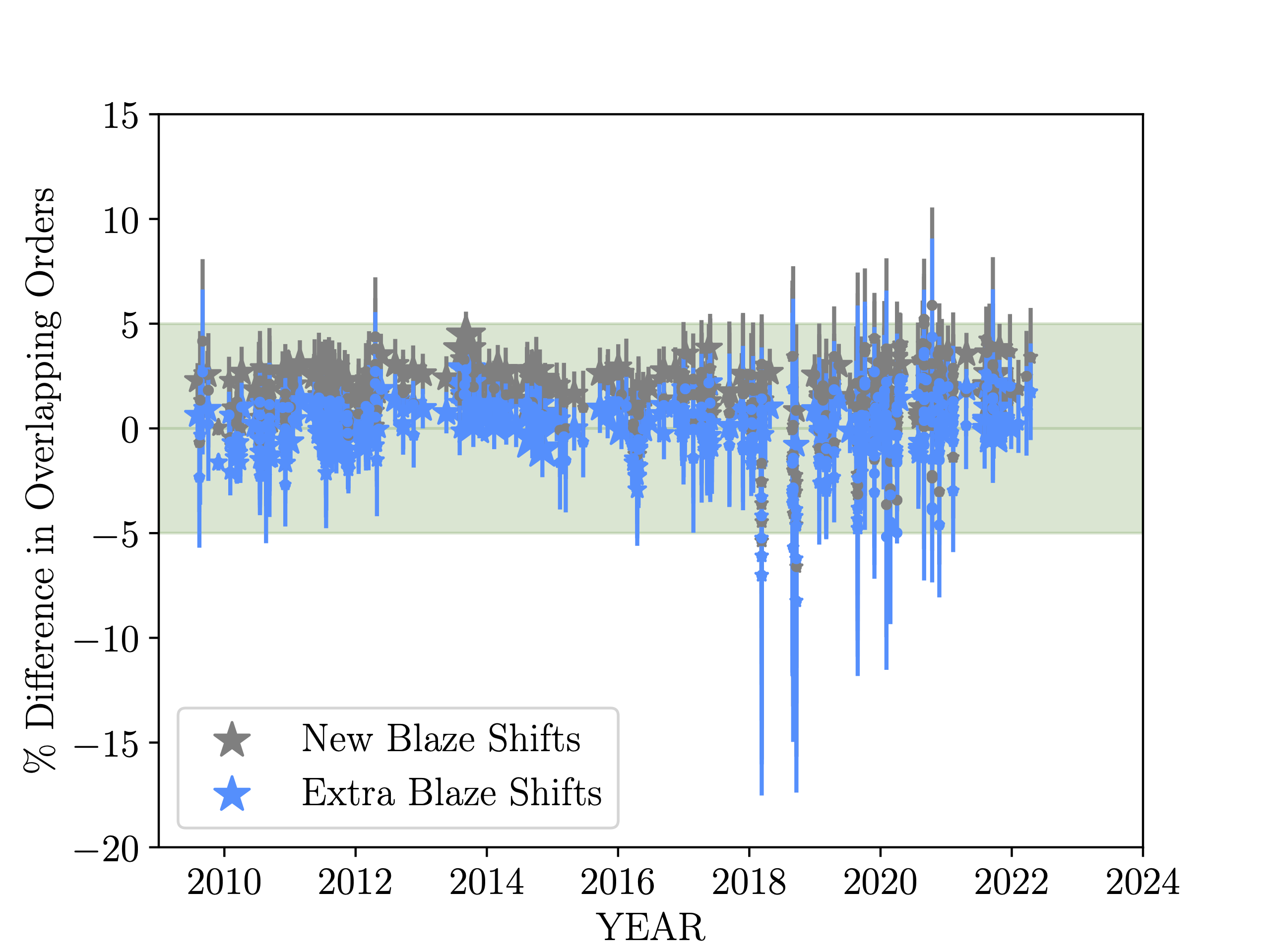} \\
    \caption{Relative flux agreement between overlapping spectral orders of post-SM4 E230M/2707 observations.  The average percent difference and RMS are represented by the star symbols and error bars, respectively, and symbol size scales with SNR.  The new time blaze coefficients were derived in the standard manner (grey points), and after removing the extra shift contribution of the reference sensitivity G 191-B2B observation, relative to the reference  \bdstar\ observation (blue points).  Separate coefficients were derived from observations taken before and after 2018.3.}
    \label{fig:e230m_2707_extra_shift}
\end{figure}

The pixel shifts computed by the \texttt{stisblazefix} tool also revealed that GO observations have a range (or scatter) of optimal blaze shifts needed at any given observation date, although there is still a linear rise in the average blaze shift trend with time.  Furthermore, reference observations of G 191-B2B for both cenwave settings have temporal blaze shifts that are a few pixels offset (green symbol in Figure \ref{fig:E230M_1978_inflection}) from the mean pixel shifts at their observation dates.  This extra discrepancy was previously unknown and had likely in the past limited the relative flux agreement of post-SM4 E230M observations by $\sim$2\%, even for early post-SM4 observations where no blaze shift time evolution had yet occurred.  Figure \ref{fig:e230m_2707_extra_shift} demonstrates the improvement in the relative flux agreement between overlapping spectral orders after removing the difference from the `extra’ blaze shift of G 191-B2B for E230M/2707 observations (blue stars).  

A modified procedure was needed to optimize the blaze shift coefficients for the secondary setting E230M/2415 as this mode is not routinely monitored with standard star observations.  In this case an empirical trial-and-error approach was adopted to optimize the \texttt{BSHIFT\_OFFSET} term using archival post-SM4 observations.  The infrequent GO usage of this mode and scatter in the pixel shifts derived from the \texttt{stisblazefix} tool at a given observation date were not conducive to deriving robust \texttt{BSHIFT\_VS\_T} and \texttt{BSHIFT\_OFFSET} coefficients independently.  Instead, we adopted the \texttt{BSHIFT\_VS\_T} slope values derived from E230M/2707 observations.  After fixing the \texttt{BSHIFT\_VS\_T} slope for the time periods before and after 2018.3, the \texttt{BSHIFT\_OFFSET} term was manually adjusted until the relative flux agreement between overlapping spectral orders was minimized using plots similar to that in Figure  \ref{fig:e230m_2707_extra_shift} to guide the adjustments.

\begin{table}[!h] 
  \centering
    \caption{Temporal blaze shift coefficients for updated E230M and E230H modes.}\label{tab:blaze_coeff_tbl}
    \def\arraystretch{1.25}
     \begin{tabular}{l c c c c}
    \hline
    \hline
   Grating & CENWAVE & REFMJD & BSHIFT\_VS\_T & BSHIFT\_OFFSET \\
    \hline
    \multicolumn{5}{c}{USEAFTER Date 2009-05-11} \\
    \hline
    E230M & 1978   & 55164.482 & 0.00201158  & -1.32565195 \\
    & 2707  &  55164.490  & 0.00204828  & -3.48637113 \\
    & 2415 &  55164.448  & 0.00204828 & 2.70394262 \\
    E230H & 2263 &  55163.485  & -0.00793479 & -4.89372739 \\
    & 2713 &  55202.645  & -0.00793479 & 0.68284026  \\
    \hline
    \multicolumn{5}{c}{USEAFTER Date 2018-03-31} \\
    \hline
    E230M & 1978  &  58233.952$^{1}$ &  0.00081563 & 2.03490850 \\
    & 2707  &  55164.490  & 0.00204787 &  -1.51387000  \\
    & 2415 &  55164.448  & 0.00204787  & 1.26352320  \\
    E230H$^{2}$ & 2263 &  55163.485  & -0.00793479 & -4.89372739 \\
    & 2713 &  55202.645  & -0.00793479 & 0.68284026  \\
    \hline
    \end{tabular}
    \\$^{1}$ The REFMJD was manually tweaked to provide a better linear fit.
    \\$^{2}$The E230H coefficients are the same for both USEAFTER dates.
\end{table}    

\begin{figure}[!]
  \centering
  \includegraphics[width=1\textwidth]{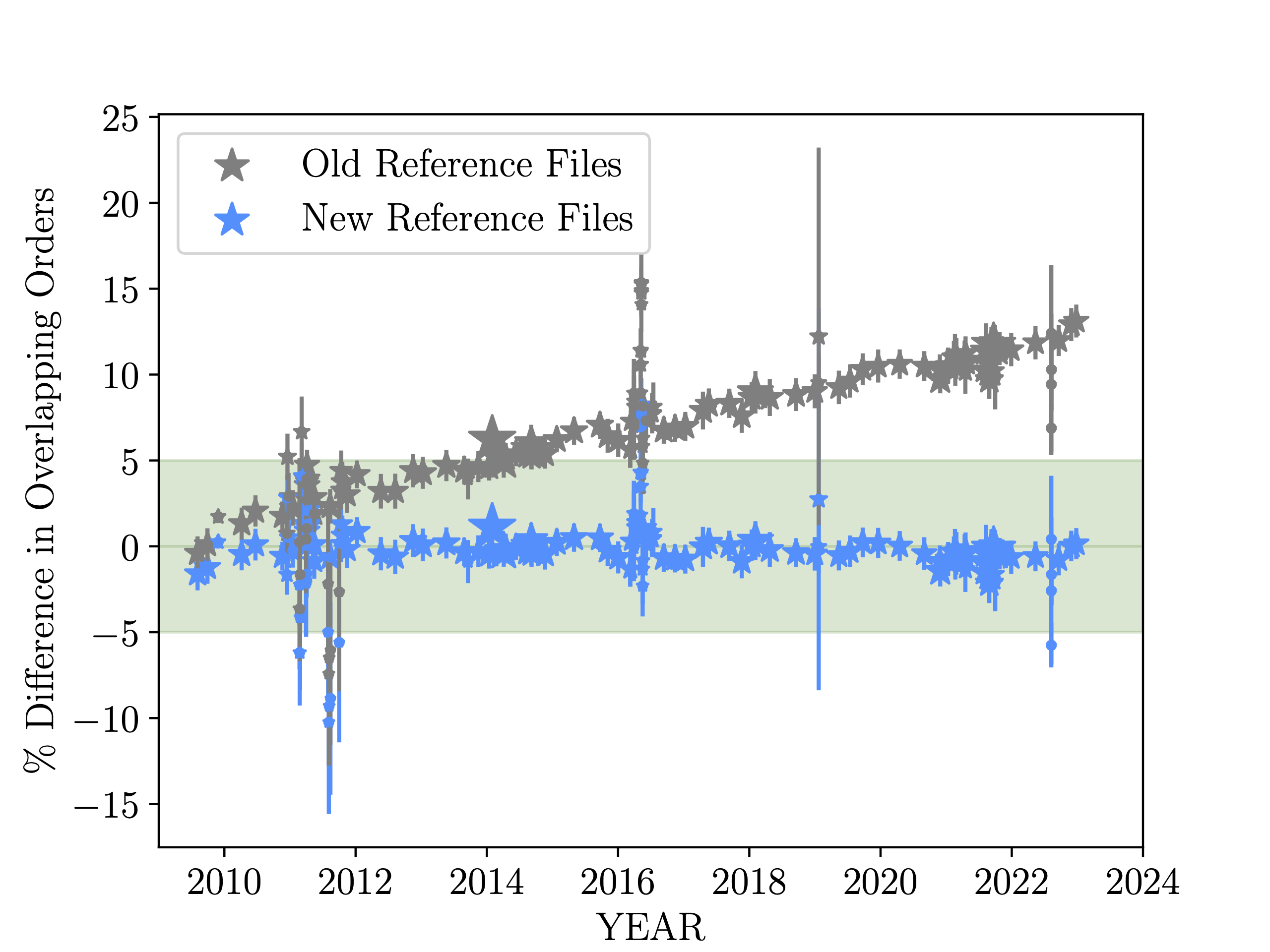} \\
    \caption{Percent differences between overlapping spectral regions of post-SM4 E230H/2263 observations of \bdstar\ using updated CALSPECv11 sensitivity curves with no blaze shifts (gray) compared to calibrations using updated sensitivity curves and newly derived blaze shift coefficients (blue).}
    \label{fig:e230h_blaze_shifts}
\end{figure}

\ssubsubsection{E230H Coefficients}\label{e230h_bs}
New blaze shift coefficients were derived for the primary setting E230H/2263 as described above using monitoring observations of \bdstar.  There was no need to remove an extra pixel shift from the G 191-B2B sensitivity curve observation for this particular setting.  We also found that a single set of coefficients was adequate to characterize all post-SM4 observations.  Figure \ref{fig:e230h_blaze_shifts} shows the excellent and much-improved relative flux agreement achieved with the new blaze shift coefficients for E230H/2263 observations. 

\begin{figure}[!]
  \centering
  \includegraphics[width=1\textwidth]{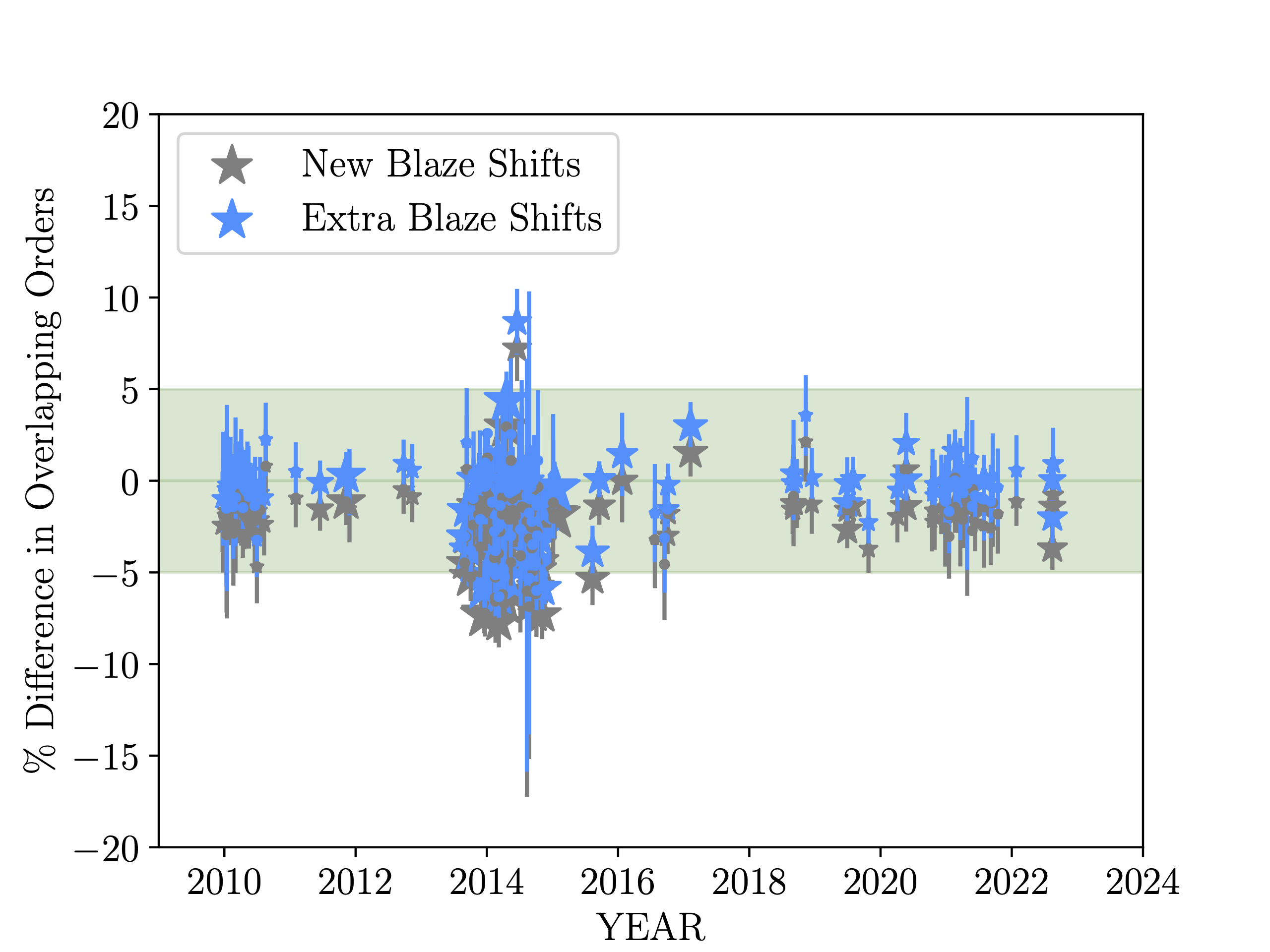} \\
    \caption{The percent differences for post-SM4 E230H/2713 observations using CALSPECv11 sensitivity curves using E230H/2713 blaze shift coefficients (gray), compared to calibrations using the updated sensitivity curves and \texttt{BSHIFT\_OFFSET} coefficients that were refined to optimize their relative flux agreement.}
    \label{fig:e230h_2713_improvement_with_time}
\end{figure}

Similar to the secondary setting E230M/2415, the E230H/2713 secondary mode is not routinely monitored with standard star observations. Due to the lack of observations, we adopted an empirical approach to optimize the coefficients.  The \texttt{BSHIFT\_VS\_T} coefficients derived for E230H/2263 were applied to the E230H/2713 setting, with an extra offset added to the \texttt{BSHIFT\_OFFSET} term and manually adjusted until the relative flux agreement between overlapping spectral regions in adjacent orders produced percent differences near zero for all of post-SM4.  Figure \ref{fig:e230h_2713_improvement_with_time} highlights the $\sim$2\% improvement achieved once the \texttt{BSHIFT\_OFFSET} coefficients were optimized for the E230H/2713 observations.

\begin{figure}[!]
  \centering
  \includegraphics[width=1\textwidth]{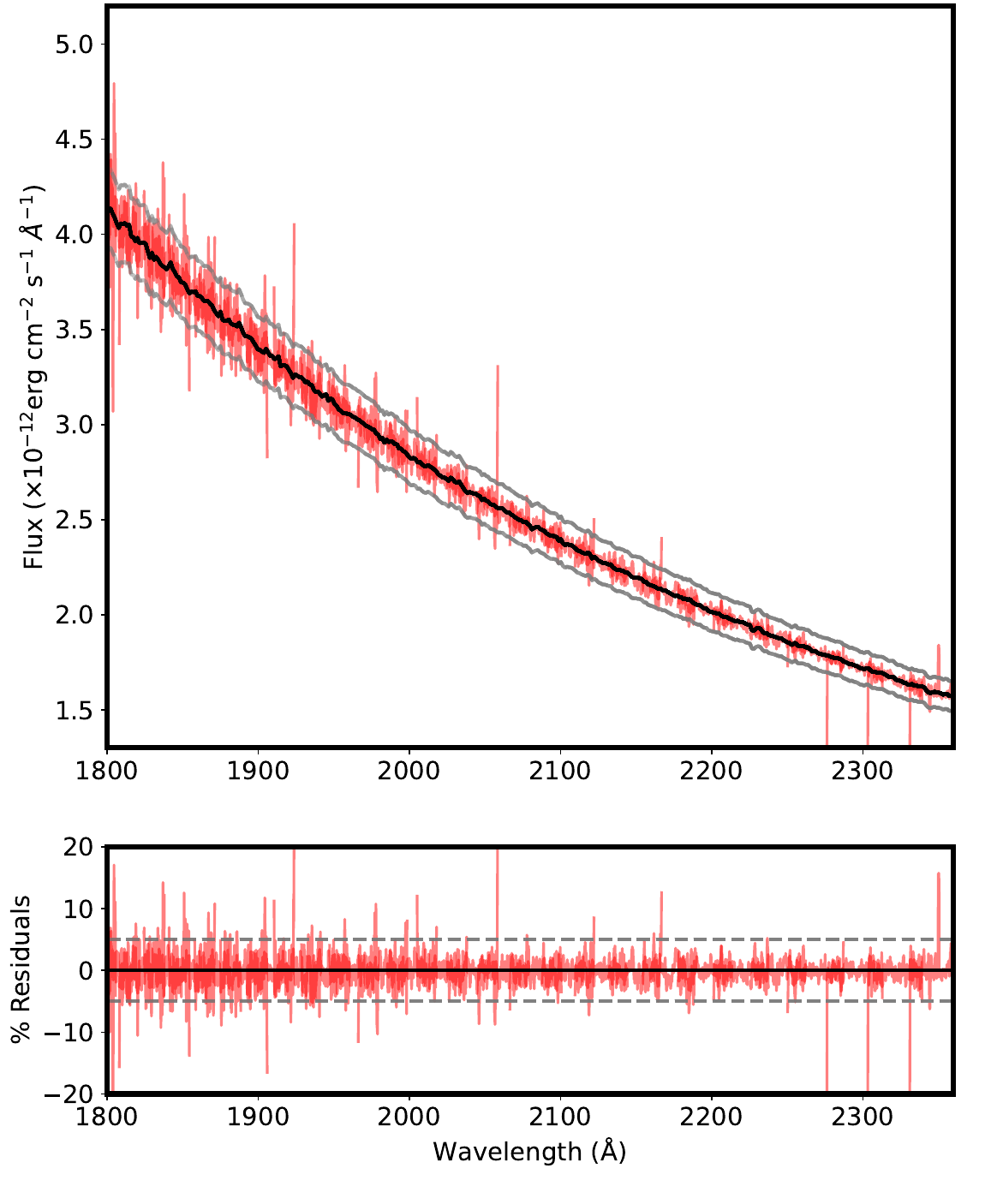} \\
    \caption{\textit{Top: }G 191-B2B spectrum taken with the E230M/1978 setting calibrated with the latest reference files in red (dataset obb004050). In black we show the STIS spectrum in the CALSPECv11 database, with the $\pm$5 \% intervals in grey. \textit{Bottom:} Percent residuals in red. The black solid line shows the 0 value, and grey dashed lines show the $\pm$5 \% intervals.}
    \label{fig:e230m_1978}
\end{figure}

\vspace{-0.3cm}
\ssection{Validation and Accuracies}\label{sec:validation}
The accuracy of the flux calibration with the new throughputs, blaze shift coefficients and ripple tables was assessed by estimating the residuals with respect to the STIS spectrum from the CALSPECv11 library. We show in Figure \ref{fig:e230m_1978} a G 191-B2B E230M/1978 spectrum calibrated with the latest reference files in red, and overplot in black the STIS spectrum in the CALSPECv11 library. We show with grey lines the $\pm$5\% flux uncertainty (see the Appendix for similar plots for the rest of the updated echelle modes). The bottom subplot in Figure \ref{fig:e230m_1978} shows the \% residual between the new echelle calibration and the STIS CALSPECv11 spectrum. We estimate the average \% residual for each order, and overall these residuals are within 1\% (with a standard deviation $\sim$1-2\% mainly driven by the S/N of the observations) for all of the updated modes. The previous flux calibration, on the other hand, showed residuals between $\sim$2-3\%, with the bluest orders in E140M/1425 exhibiting the largest residuals. We highlight that the improvements in the agreement between the previous and new calibrations against the continuum in the new CALSPECv11 models is a combination of both the re-derivation of the echelle sensitivities as well as the updates to the blaze shift coefficients (e.g., Figure \ref{fig:blaze_e230m_2707}).\par
According to \href{https://ui.adsabs.harvard.edu/abs/2020AJ....160...21B/abstract}{Bohlin et al. (2020)}, the most recent SEDs produced by the NLTE model atmosphere codes for G 191-B2B have shown systematic differences in the FUV of the order of $\le$4\%, and in the NUV of the order of $\sim$1.5\%. Accounting for instrumental uncertainties (e.g., time-dependent sensitivity, focus-dependent throughput) of the order of $\sim$4\% for both FUV and NUV (\href{https://www.stsci.edu/files/live/sites/www/files/home/hst/instrumentation/stis/documentation/instrument-science-reports/_documents/2017_06.pdf}{Carlberg et al. 2017}), we estimate absolute flux calibration accuracies of 6\% for the FUV, and 4\% for the NUV coverage. \par

Lastly, we note that similar to the work by \href{https://www.stsci.edu/files/live/sites/www/files/home/hst/instrumentation/stis/documentation/instrument-science-reports/_documents/2012_01.pdf}{Bostroem et al. (2012; their Section 2.1)}, as part of our testing we calibrated all of the existing TDS monitoring data using the updated reference files to assess if any of the observations used to update the sensitivities were impacted by anomalously low throughputs (known as flux anomalies; \href{https://www.stsci.edu/files/live/sites/www/files/home/hst/instrumentation/stis/documentation/instrument-science-reports/_documents/2012_01.pdf}{Bostroem et al. 2012}). We compared the newly calibrated fluxes against the corresponding CALSPECv11 spectrum, and overall we observed flux variations of the order of $\lesssim$5\%, well within the expected absolute photometric accuracy for the echelle modes (\href{https://hst-docs.stsci.edu/stisihb/chapter-16-accuracies/16-1-summary-of-accuracies}{STIS Instrument Handbook Chapter 16}).

\vspace{-0.3cm}
\ssection{Summary}\label{sec:summary}
\vspace{-0.3cm}
In this report we presented a general overview of the steps taken to update the throughputs, ripple functions and blaze shift coefficients for a list of prioritized echelle modes, specifically for post-SM4 observations. These updates were primarily driven by the flux changes introduced by the primary standards in the CALSPECv11 library (\href{https://ui.adsabs.harvard.edu/abs/2020AJ....160...21B/abstract}{Bohlin et al. 2020}), over the CALSPECv07 SEDs previously used in the throughout derivation of the STIS echelle modes (\href{https://www.stsci.edu/files/live/sites/www/files/home/hst/instrumentation/stis/documentation/instrument-science-reports/_documents/2012_01.pdf}{Bostroem et al. 2012}). The updated CALSPECv11 models increase the UV fluxes by $\sim$2-3\%.

\newpage



\vspace{-0.3cm}
\ssectionstar{Acknowledgements}
\vspace{-0.3cm}
We thank Ralph Bohlin for valuable discussions on flux calibrations with STIS.

\vspace{-0.3cm}
\ssectionstar{Change History for STIS ISR 2024-04}\label{sec:History}
\vspace{-0.3cm}
Version 1: 21 August 2024 - Original Document 

\vspace{-0.3cm}
\ssectionstar{References}\label{sec:References}
\vspace{-0.3cm}

\noindent
Aloisi, A., Bohlin, R., \& Kim Quijano, J. 2007, STIS Instrument Science Report 2007- 01\\
Bohlin, R., Hubeny, I., \& Rauch, T. 2020, AJ, 160, 21\\
Bostroem, K. A., Aloisi, A., Bohlin, R., Hodge, P., \& Proffitt, C. 2012, STIS Instrument Science Report 2012-01\\
Carlberg, J., Monroe, T., Riley, A., \& Hernandez, S., 2022, STIS Instrument Science Report 2022-04\\
Carlberg, J., \& Monroe, T., 2017, STIS Instrument Science Report 2017-06\\
Medallon, S., Rickman, E., Brown, J., 2023 “STIS Instrument Handbook (IHB),” Version 23.0, (Baltimore: STScI)\\
Reid, N. \& Wegner, G. 1988, ApJ, 335,953\\
Siebert, M. R., Carlberg, J. K., Hernandez, S., Monroe, T., 2024, STIS Instrument Science Report 2024-02\\  
Valenti, J., Lindler, D., Bowers, C., Busko, I., \& Kim Quijano, J., 2002, STIS Instrument Science Report 2002-01\\

\vspace{-0.3cm}
\ssectionstar{Appendix A}\label{sec:Appendix}
\vspace{-0.3cm}
G 191-B2B spectra comparing the latest echelle calibration against the CALSPECv11 model. 

\begin{figure}[!]
  \centering
  \includegraphics[width=1\textwidth]{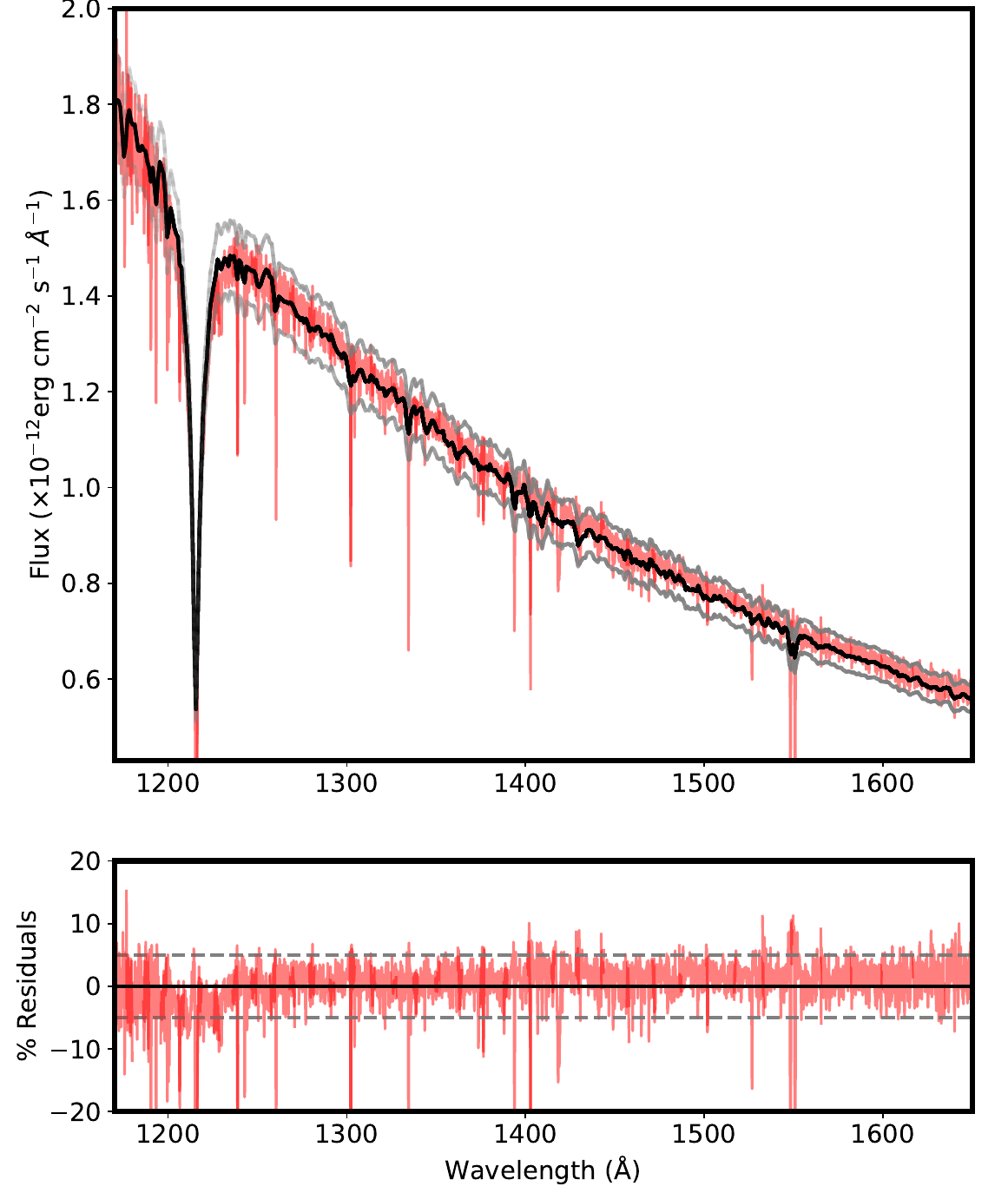} \\
    \caption{\textit{Top: }G 191-B2B spectrum taken with the E140M/1425 setting calibrated with the latest reference files in red. In black we show the STIS spectrum in the CALSPECv11 database, with the $\pm$5 \% intervals in grey. \textit{Bottom:} Percent residuals in red. The black solid line shows the 0 value, and grey dashed lines show the $\pm$5 \% intervals.}
    \label{fig:e140m_1425}
\end{figure}

\begin{figure}[!]
  \centering
  \includegraphics[width=1\textwidth]{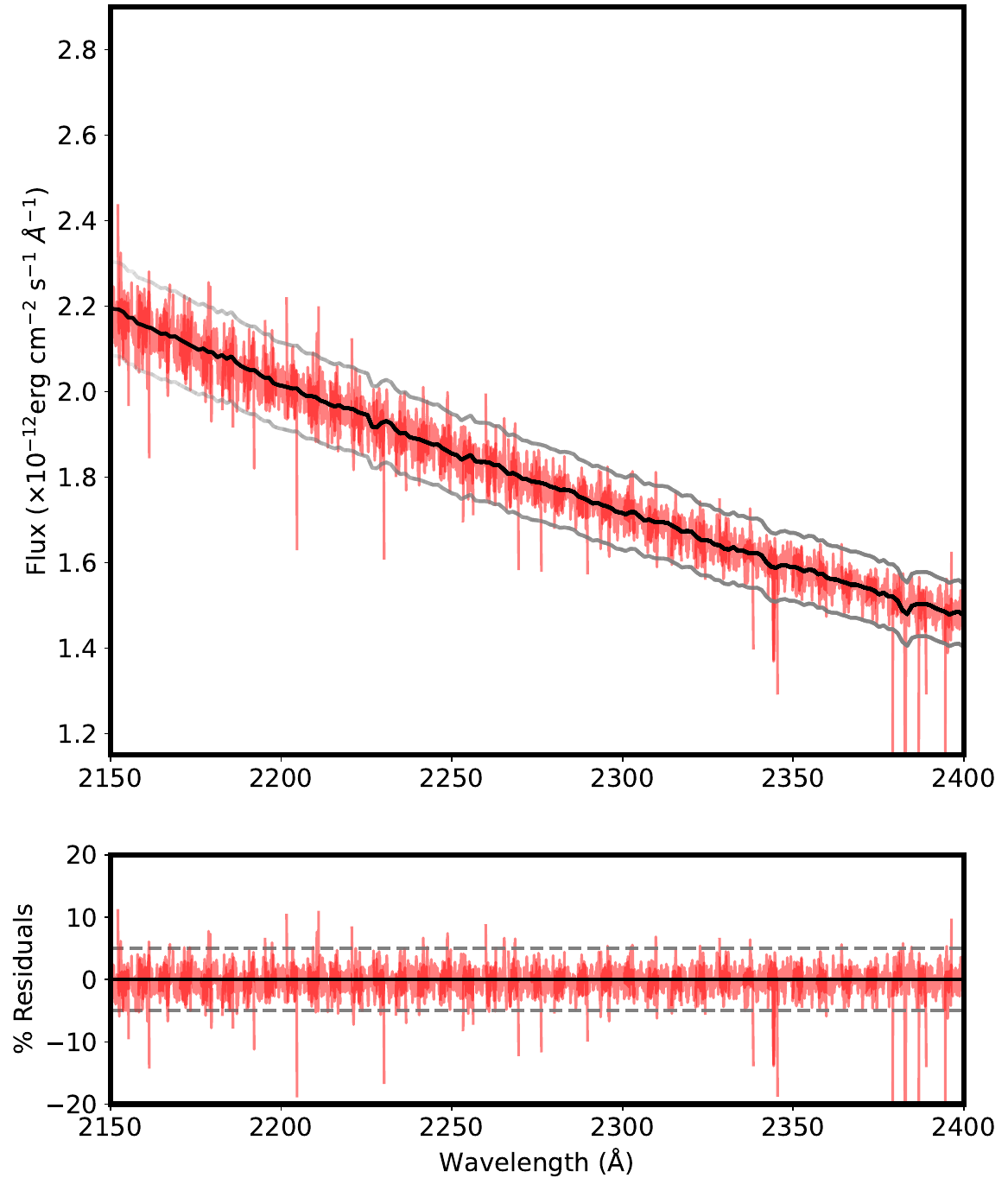} \\
    \caption{\textit{Top: }G 191-B2B spectrum taken with the E230H/2263 setting calibrated with the latest reference files in red. In black we show the STIS spectrum in the CALSPECv11 database, with the $\pm$5 \% intervals in grey. \textit{Bottom:} Percent residuals in red. The black solid line shows the 0 value, and grey dashed lines show the $\pm$5 \% intervals.}
    \label{fig:e230h_2263}
\end{figure}

\begin{figure}[!]
  \centering
  \includegraphics[width=1\textwidth]{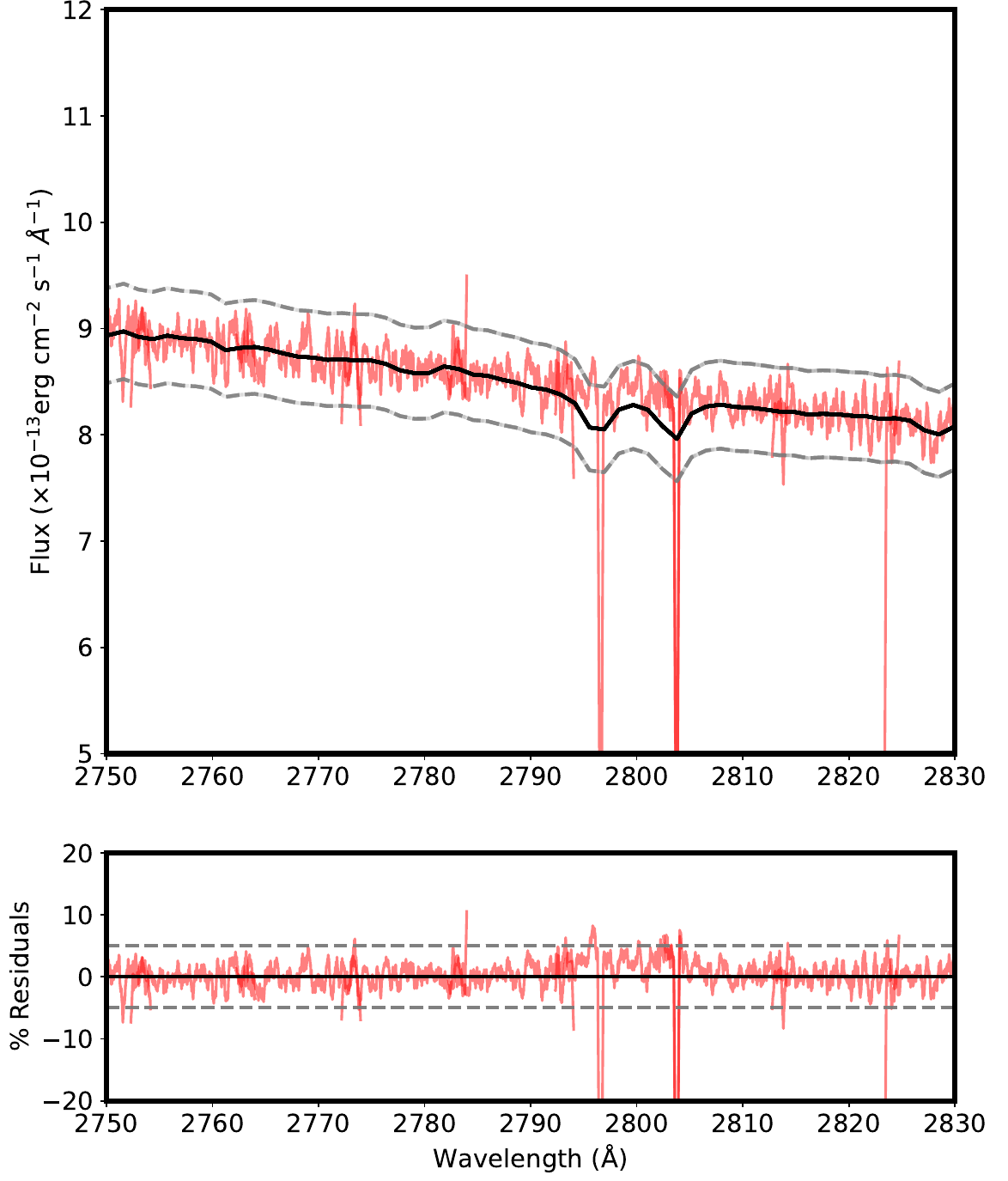} \\
    \caption{\textit{Top: }G 191-B2B spectrum taken with the E230H/2713 setting calibrated with the latest reference files in red. In black we show the STIS spectrum in the CALSPECv11 database, with the $\pm$5 \% intervals in grey. \textit{Bottom:} Percent residuals in red. The black solid line shows the 0 value, and grey dashed lines show the $\pm$5 \% intervals.}
    \label{fig:e230h_2713}
\end{figure}

\begin{figure}[!]
  \centering
  \includegraphics[width=1\textwidth]{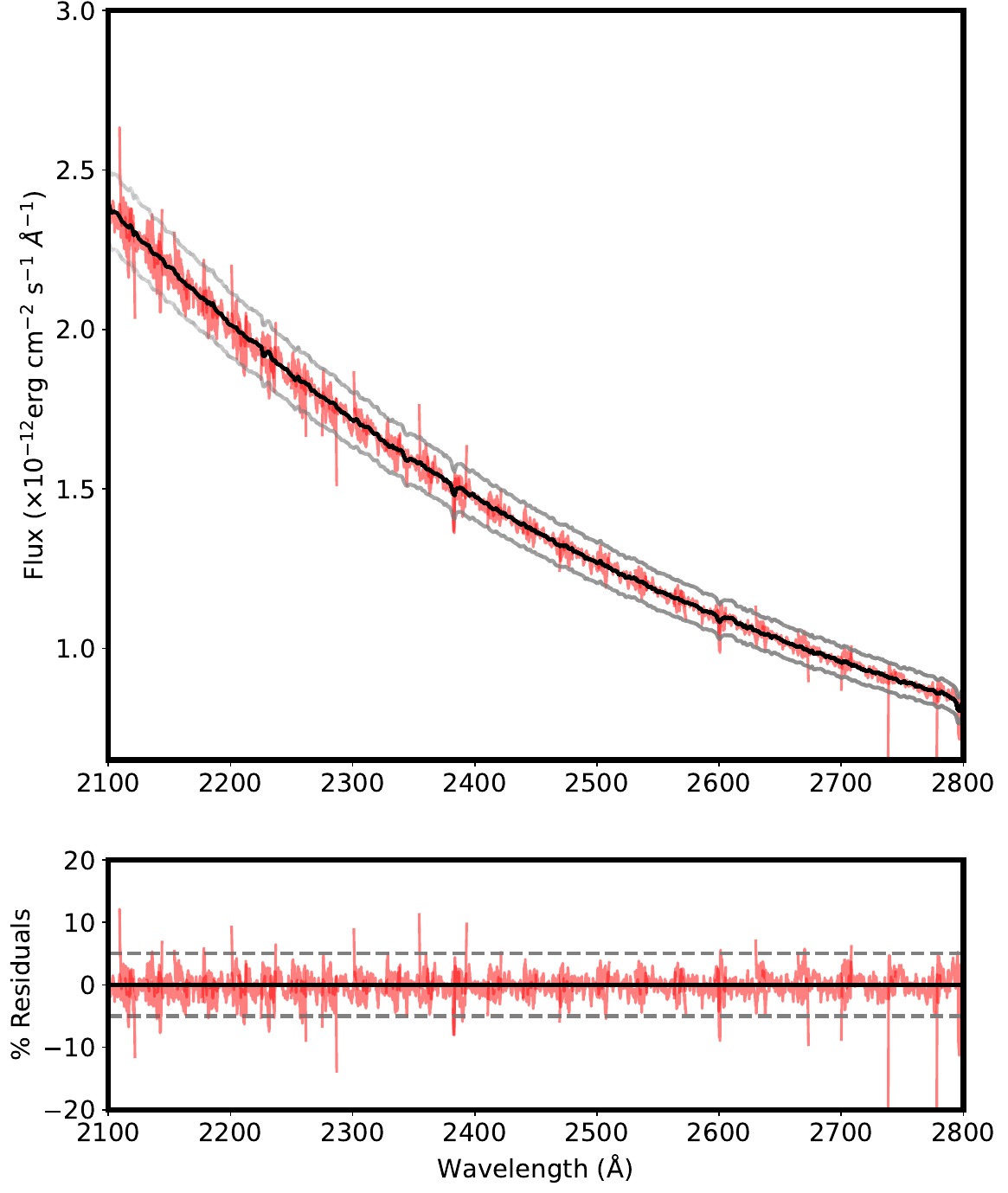} \\
    \caption{\textit{Top: }G 191-B2B spectrum taken with the E230M/2415 setting calibrated with the latest reference files in red. In black we show the STIS spectrum in the CALSPECv11 database, with the $\pm$5 \% intervals in grey. \textit{Bottom:} Percent residuals in red. The black solid line shows the 0 value, and grey dashed lines show the $\pm$5 \% intervals.}
    \label{fig:e230m_2415}
\end{figure}

\begin{figure}[!]
  \centering
  \includegraphics[width=1\textwidth]{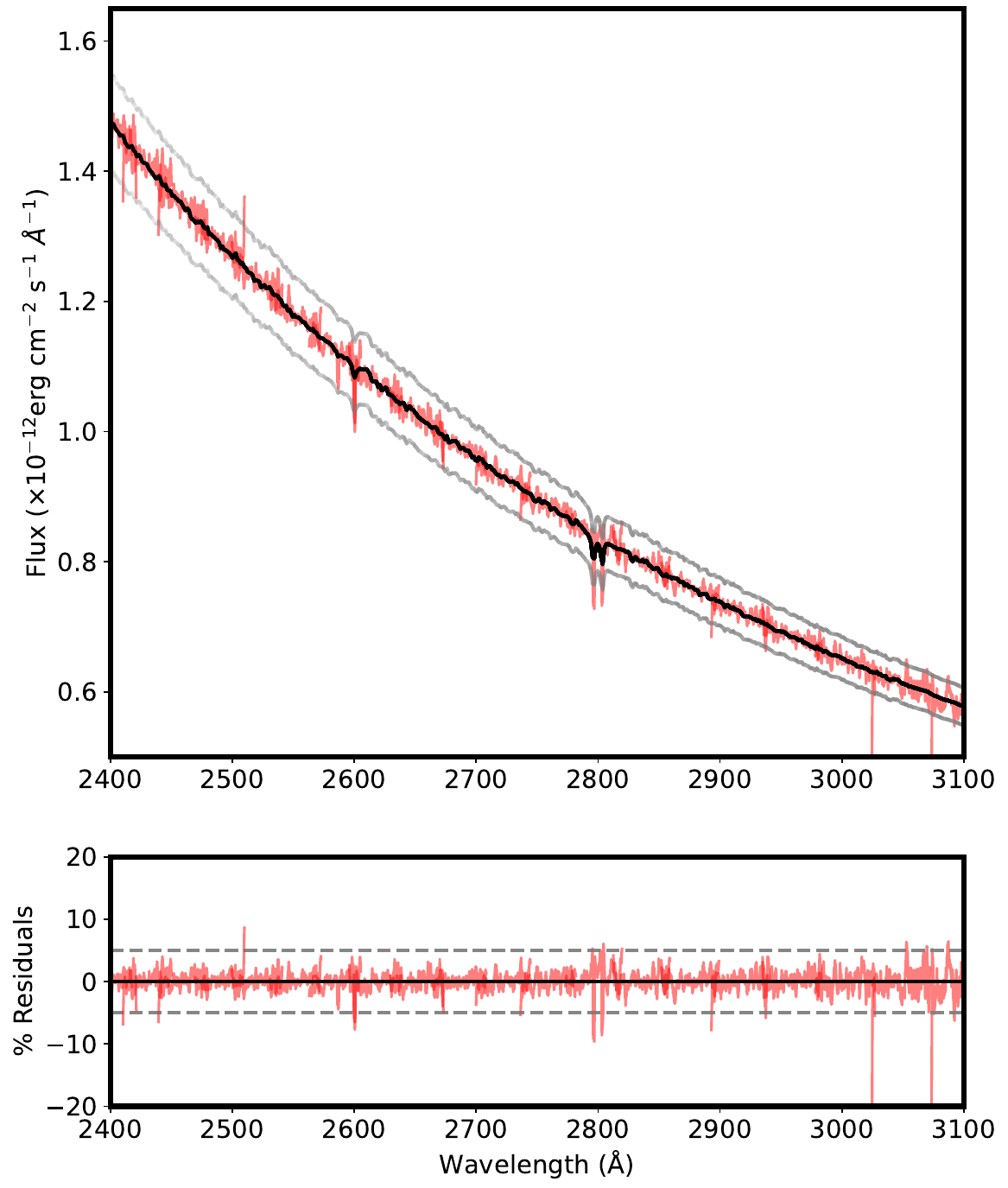} \\
    \caption{\textit{Top: }G 191-B2B spectrum taken with the E230M/2707 setting calibrated with the latest reference files in red. In black we show the STIS spectrum in the CALSPECv11 database, with the $\pm$5 \% intervals in grey. \textit{Bottom:} Percent residuals in red. The black solid line shows the 0 value, and grey dashed lines show the $\pm$5 \% intervals.}
    \label{fig:e230m_2707}
\end{figure}

\end{document}